\documentclass[12pt,twoside]{article}
\usepackage{epsfig}
\usepackage{graphicx}
\usepackage{amssymb}
\title{Electromagnetic Boundary Conditions\\ Defined in Terms of\\ Normal Field Components}
\author{I.V. Lindell and A.H. Sihvola} 
\date{Department of Radio Science and Engineering\\ Helsinki University of Technology\\ Box 3000, Espoo 02015TKK, Finland\\ {\tt ismo.lindell@tkk.fi}}
\def\e{\begin{equation}} 
\def\f{\end{equation}} 

\def\##1{{\bf #1\mit}}
\def\=#1{{\overline{\overline{\mathsf #1}}}}
\def\RR{\mbox{\boldmath $\R$}}

\def\*{^{\displaystyle*}}
\def\.{\cdot}
\def\x{\times}
\def\oo{\infty}

\def\D{\nabla}
\def\d{\partial}

\def\ra{\rightarrow}

\def\Ra{\Rightarrow}
\def\le{\left(}
\def\ri{\right)}
\def\l#1{\label{eq:#1}}
\def\r#1{(\ref{eq:#1})}
\def\am{\left(\begin{array}{c}}
\def\amm{\left(\begin{array}{cc}}
\def\ammm{\left(\begin{array}{ccc}}
\def\ammmm{\left(\begin{array}{cccc}}
\def\a{\end{array}\right)}

\def\add{\left|\begin{array}{cc}}
\def\addd{\left|\begin{array}{ccc}}
\def\adddd{\left|\begin{array}{cccc}}
\def\ad{\end{array}\right|}

\def\I{\int\limits}
\def\OI{\oint\limits}

\def\E{\epsilon}

\def\h{\eta}

\def\M{\mu}
\def\o{\omega}

\def\R{\rho}

\def\TH{\theta}

\def\VF{\varphi}

\begin{document}

\maketitle

\begin{abstract}
A set of four scalar conditions involving normal components of the fields $\#D$ and $\#B$ and their normal derivatives at a planar surface is introduced, among which different pairs can be chosen to represent possible boundary conditions for the electromagnetic fields. Four such pairs turn out to yield meaningful boundary conditions and their responses for an incident plane wave at a planar boundary are studied. The theory is subsequently generalized to more general boundary surfaces defined by a coordinate function. It is found that two of the pairs correspond to the PEC and PMC conditions while the other two correspond to a mixture of PEC and PMC conditions for fields polarized TE or TM with respect to the coordinate defining the surface. 
\end{abstract}

\section{Introduction}

When representing electromagnetic field problems as boundary-value problems the boundary conditions are generally defined in terms of fields tangential to the boundary surface. A typical example is that of impedance boundary conditions which can be expressed in the form
\cite{Pelosi,Methods}
\e \#n\x(\#E - \=Z_s\.(\#n\x\#H))=0,\ \ \ \ \#n\.\=Z_s=\=Z_s\.\#n=0, \l{imp} \f
for some surface impedance dyadic $\=Z_s$ which may have infinite components. Here, $\#n$ is the unit vector normal to the boundary surface. Special cases for the impedance boundary are the perfect electric counductor (PEC) boundary, 
\e \=Z_s=0\ \ \ \ \Ra\ \ \ \ \#n\x\#E=0, \l{PEC}\f
and the perfect magnetic conductor (PMC) boundary \cite{Monorchio,Feresidis},
\e \=Z_s\ra \oo,\ \ \ \ \Ra\ \ \ \ \#n\x\#E=0, \l{PMC}\f
also known as high-impedance surface \cite{Yablonovich, Clavijo,Luukkonen}.
A generalization of these is the perfect electromagnetic conductor (PEMC) boundary defined by
\e \=Z_s = \frac{1}{M}\#n\x\=I,\ \ \ \ \Ra\ \ \ \ \#n\x(M\#E+\#H)=0, \l{PEMC}\f
where $M$ is the PEMC admittance \cite{PEMC}. In fact, for $M=0$ and $1/M=0$ \r{PEMC} yields the respective PMC and PEC boundaries.

A different set of boundary conditions involving normal components of the vectors $\#D$ and $\#B$ at the boundary surface,
\e \#n\.\#D=0,\ \ \ \ \ \#n\.\#B=0  \l{DB}\f
has been recently introduced in \cite{Kong,Yaghjian} in conjunction with electromagnetic cloaking and, independently, by these authors \cite{URSI08,Hongkong,IBDB,DB}. A boundary defined by the conditions \r{DB} was dubbed {\it DB boundary} for brevity. The corresponding conditions for the $\#E$ and $\#H$ vectors depend on the medium in front of the boundary. In this study we assume a simple isotropic medium with permittivity $\E$ and permeability $\M$, whence \r{DB} is equivalent to the conditions
\e \#n\.\#E=0,\ \ \ \ \ \#n\.\#H=0.  \l{EH}\f
A more recent study of literature has revealed that the DB conditions either in the form of \r{DB} or \r{EH} have been considered much earlier. In fact, in 1959 V.H. Rumsey discussed the uniqueness of a problem involving the conditions \r{EH} as well as their realization in terms of the interface of a uniaxially anisotropic medium \cite{Rumsey}. The uniqueness and existence problems were considered more exactly in subsequent papers \cite{Yee,Picard74,Picard77,Kress,Gulzow}. 
 
The DB-boundary conditions \r{DB} were introduced by these authors in \cite{IBDB} as following from the interface conditions for the half space $z<0$ of a certain exotic material called uniaxial IB or skewon-axion medium \cite{IB,Hehl}. A more general proof is given in Appendix 1. A simpler realization for the DB boundary appears possible in terms of the planar interface $z=0$ of a uniaxially anisotropic medium defined by the permittivity and permeability dyadics \cite{Rumsey,IBDB}
\e \=\E = \E_z\#u_z\#u_z + \E_t\=I_t,\ \ \ \ \=\M= \M_z\#u_z\#u_z + \M_t\=I_t, \l{uniax}\f
with the transverse unit dyadic defined by 
\e \=I_t = \=I-\#u_z\#u_z = \#u_x\#u_x + \#u_y\#u_y. \f
Assuming vanishing axial parameters, $\E_z\ra0,\ \M_z\ra0$, the fields in the uniaxial medium $z<0$ satisfy $D_z\ra0$ and $B_z\ra0$, whence because of the continuity, the DB conditions \r{DB} are valid at the interface $z=0$. Such a uniaxial medium was dubbed zero axial parameter (ZAP) medium \cite{ZAP}. Obviously, the same principle applies for curved boundaries as well, when the medium is locally uniaxial with vanishing normal components of permittivity and permeability. Zero-valued electromagnetic parameters and their applications have been studied recently together with their realizations in terms of metamaterials \cite{Engheta,Silveirinha,Alu,Luukkonen1}.

Basic properties of the DB boundary for electromagnetic fields have been recently studied. In \cite{IBDB} it was shown that, since the Poynting vector has only the normal component, the DB boundary is an isotropic soft surface in the definition of Kildal \cite{SHS,Kildal}. Thus, coupling between aperture antennas on a DB plane is smaller than on a PEC plane. Further properties of a planar DB boundary were analyzed in \cite{DB}. It was shown that the DB plane can be replaced by a PEC plane for fields polarized TE$^z$ with respect to the normal ($z$) direction and, by a PMC plane for TM$^z$ fields. Thus, radiation from a current source $\#J$ in front of the DB plane can be found by splitting the source in two parts, $\#J_{\rm TE}$ radiating a TE$^z$ field and $\#J_{\rm TM}$ radiating a TM$^z$ field, whence the DB plane can be replaced by the images of the two source components, PEC image for the TE$^z$ component and PMC image for the TM$^z$ component. 

In \cite{EM5} the resonator defined by a spherical DB boundary was studied. Splitting all modes in two sets, those polarized TE$^r$ and TM$^r$ with respect to the radial coordinate $r$, it was shown that the TE$^r$ and the TM$^r$ modes equal those of the respective PEC and PMC resonator. Thus, the number of modes with the same resonance frequency is double of that of the PEC or PMC resonator which means that there is more freedom to define the resonance field. 

Finally, the circular waveguide defined by the DB boundary was analyzed in \cite{EM7}. Since general modes cannot be decomposed in TE$^\R$ and TM$^\R$ parts with respect to the polar radial coordinate $\R$, the modes had to be computed in the classical way. It was shown that there may exist backward-wave modes  in such a waveguide although there is no dispersion or periodic structure involved. 

The objective of the present paper is to extend the concept of DB boundary by adding another set of conditions involving normal derivatives of the normal field components. First, basic properties of the plane-wave reflected from planar boundaries satisfying different boundary conditions are derived, after which the more general curved boundary defined by a coordinate function will be studied.

\section{Planar boundary conditions}

In the following we consider a planar boundary defined by $z=0$ and fields in the half space $z\geq0$. The DB condition \r{DB} can be expressed as vanishing of the $z$ components of the two fields,
\e D_z=0,\ \ \ \ \ B_z=0, \l{DBz}\f
at the boundary. If there are no sources at the boundary, the $z$ components of the Maxwell equations yield conditions for the transverse components of the fields as 
\e \D_t\x\#H_t=0,\ \ \ \ \ \D_t\x\#E_t=0. \l{Dtx}\f
Obviously, the conditions \r{Dtx} are equivalent to \r{DBz}.

Let us now introduce another possible set of boundary conditions involving the normal derivatives of the field components at the planar surface:
\e \d_zD_z=0,\ \ \ \ \ \d_zB_z=0, \l{DzBz}\f
and let us call such a boundary by the name {\it D'B' boundary} for brevity. If there are no sources at the D'B' boundary, the fields $\#D$ and $\#B$ are solenoidal and they satisfy
\e \D\.\#D=\d_zD_z + \D_t\.\#D_t =0,\ \ \ \ \D\.\#B=\d_zB_z+\D_t\.\#B_t=0. \f
Thus, the D'B' conditions \r{DzBz} can be alternatively expressed in terms of the transverse components of the fields as
\e \D_t\.\#D_t=0,\ \ \ \ \D_t\.\#B_t=0. \l{DtBt}\f
In an isotropic medium the D'B' conditions \r{DzBz} and \r{DtBt} are equivalent to
\e \d_zE_z=0,\ \ \ \ \ \d_zH_z=0, \l{EzHz}\f
\e \D_t\.\#E_t=0,\ \ \ \ \D_t\.\#H_t=0. \l{EtHt}\f

There are two other combinations of the four conditions in \r{DBz} and \r{DzBz} which may appear useful. According to the previous pattern, let us call the conditions
\e D_z=0,\ \ \ \ \ \d_zB_z=0 \l{DB'}\f
as those of the DB' boundary and the conditions
\e \d_zD_z=0,\ \ \ \ \ B_z=0 \l{D'B}\f
as those of the D'B boundary. In an isotropic medium \r{DB'} can be replaced by
\e \D_t\x\#H_t=0,\ \ \ \ \D_t\.\#H_t=0, \l{DB'1} \f
while the conditions \r{D'B} can be replaced by
\e \D_t\.\#E_t=0,\ \ \ \ \D_t\x\#E_t=0. \l{D'B1} \f
From \r{DB'1} it follows that there exists a scalar potential $\psi(x,y)$ in terms of which we can express the field $\#H_t$ at the DB' boundary and $\psi$ satisfies the Laplace equation:
\e \#H_t(x,y) = \D_t\psi(x,y),\ \ \ \ \D_t^2\psi(x,y) = 0. \l{Hpsi}\f
Similarly, we can write for the field $\#E_t$ at the D'B boundary in terms of a potential $\phi$ as
\e \#E_t(x,y) = \D_t\phi(x,y),\ \ \ \ \D_t^2\phi(x,y) = 0. \l{Ephi}\f
Assuming a localized source, the tangential components of the radiation fields are known to decay in the infinity as $1/r$ and the radial components a $1/r^2$ \cite{Bladel}. Following the argumentation of the Appendix, we must then have $\#H_t=0$ and $\#E_t=0$ at the boundary surface. Thus, under the assumption of localized sources the DB' conditions equal the PMC condition \r{PMC} and the D'B conditions equal the PEC condition \r{PEC}. However, this is not valid for all non-localized sources. For example, a constant planar surface current gives rise to a TEM field with $D_z=0$ and $B_z=0$ everywhere, whence \r{DB'} and \r{D'B} are satisfied identically. This effect was discussed for the DB boundary in \cite{DB,ZAP}.

\section{Reflection of plane wave}

\subsection{Plane-wave relations}

The basic problem associated with the D'B' boundary is to find the reflection of a plane wave from the planar boundary $z=0$ in the isotropic half space $z>0$ defined by the parameters $\E,\M$. Assuming $\exp(j\o t)$ time dependence and choosing the $x$ axis so that the wave vector of the incident and reflected waves are in the $xz$ plane, the field and wave vectors have the form 
\e \#E^i(\#r) = \#E^ie^{-j\#k^i\.\#r},\ \ \ \ \#H^i(\#r) = \#H^ie^{-j\#k^i\.\#r}, \f  
\e \#E^r(\#r) = \#E^re^{-j\#k^r\.\#r},\ \ \ \ \#H^r(\#r) = \#H^re^{-j\#k^r\.\#r}, \f  
\e \#k^i= \#u_xk_x -\#u_z k_z,\ \ \ \ \#k^r=\#u_x k_x + \#u_z k_z, \f
Because the fields of a plane wave are divergenceless, they satisfy the orthogonality relations
\e \#k^i\.\#E^i=0,\ \ \ \#k^i\.\#H^i=0, \f
\e \#k^r\.\#E^r=0,\ \ \ \#k^r\.\#H^r=0. \f
Assuming $k_x\not=0$, i.e., excluding the normal incidence case, the fields can be expressed in terms of their $z$ components as
\e \#E^i = (\#u_z + \#u_x k_z/k_x)E_z^i + \#u_y(k/k_x)\h H_z^i, \l{Ei}\f
\e \h\#H^i =(\#u_z + \#u_x k_z/k_x)\h H_z^i -\#u_y(k/k_x)E_z^i, \l{Hi}\f
\e \#E^r = (\#u_z - \#u_x k_z/k_x)E_z^r + \#u_y(k/k_x)\h H_z^r, \l{Er}\f
\e \h\#H^r = (\#u_z - \#u_x k_z/k_x)\h H_z^r - \#u_y(k/k_x)E_z^r. \l{Hr}\f
Here, 
\e k=\o\sqrt{\M\E},\ \ \ \ \h = \sqrt{\M/\E} \f
denote the respective wavenumber and wave-impedance quantities. 

Let us consider an incident plane wave whose $z$ components satisfy the condition
\e E_z^i\sin\TH+  \h H_z^i\cos\TH =0. \l{TH}\f
Actually, any plane wave satisfies a condition of the form \r{TH} for some parameter $\TH$. (The TEM case $E_z^i=0$, $H_z^i=0$ is excluded because of the assumption $k_x\not=0$.) The special cases of TE and TM waves correspond to the respective cases $\cos\TH=0$ and $\sin\TH=0$. From \r{Ei}, \r{Hi} the incident field vectors tangential to the boundary can be shown to satisfy the two conditions
\e \#E^i_t\sin\TH + \h \#H^i_t\cos\TH =\#u_y\frac{k}{k_x}(\h H_z^i\sin\TH- E_z^i\cos\TH), \l{EiTH}\f
\e \#E^i_t\cos\TH -\h \#H^i_t\sin\TH =-\#u_x\frac{k_z}{k_x}(\h H_z^i\sin\TH- E_z^i\cos\TH). \l{EiTH1}\f

\subsection{DB boundary}

The DB boundary conditions \r{DBz} at $z=0$ require
\e E_z^r +E_z^i=0,\ \ \ \ \ H_z^r+ H_z^i=0, \f
whence the reflected field components $E_z^r,H_z^r$ satisfy a relation of the same form \r{TH} as the incident field components $E_z^i,H_z^i$.

From \r{Er}, \r{Hr} we see that the tangential components of the reflected field vectors satisfy
\e \#E^r_t\sin\TH + \h \#H^r_t\cos\TH =-\#u_y\frac{k}{k_x}(\h H_z^i\sin\TH- E_z^i\cos\TH). \l{ErTH}\f
Combining this with \r{EiTH} the condition for the total tangential field at the boundary becomes
\e \#E_t\sin\TH+  \h \#H_t\cos\TH =0, \l{THDB}\f
which is of the same form \r{TH} as for the $z$ components of the incident field. 

This leads us to the following conclusions:
\begin{itemize}
\item For the TE incident wave with $\cos\TH=0$, \r{THDB} yields $\#E_t=0$ which corresponds to the PEC condition \r{PEC}.
\item For the TM incident wave with $\sin\TH=0$, \r{THDB} yields $\#H_t=0$ which corresponds to the PMC condition \r{PMC}.
\item Denoting $M =\tan\TH/\h$, \r{THDB} yields $M\#E_t+\#H_t=0$  which corresponds to the PEMC condition \r{PEMC}.
\end{itemize}

\subsection{D'B' boundary}

At the D'B' boundary $z=0$ the fields satisfying the conditions \r{EzHz} yield
\e \d_z(E^i_ze^{j k_z z}+E_z^re^{-j k_z z})_{z=0}=jk_z(E^i_z-E^r_z)=0,\f
\e \d_z(H_z^ie^{j k_z z}+H_z^re^{-j k_z z})_{z=0}=jk_z(H^i_z-H^r_z)=0. \f
whence, again, the reflected field components $E_z^r,H_z^r$ satisfy a relation of the same form \r{TH} as the incident field components $E_z^i,H_z^i$.

From \r{Er}, \r{Hr} we now see that the tangential components of the reflected field vectors satisfy
\e \#E^r_t\cos\TH - \h \#H^r_t\sin\TH =-\#u_x\frac{k_z}{k_x}(\h H_z^i\sin\TH- E_z^i\cos\TH). \l{ErTH'}\f
Combining with \r{EiTH1} the condition for the total tangential fields at the boundary becomes
\e \#E_t\cos\TH - \h \#H_t\sin\TH =0. \l{THD'B'}\f

This condition leads us to the following conclusions:
\begin{itemize}
\item For the TE incident wave with $\cos\TH=0$, \r{THD'B'} yields $\#H_t=0$ which corresponds to the PMC condition \r{PMC}.
\item For the TM incident wave with $\sin\TH=0$, \r{THD'B'} yields $\#E_t=0$ which corresponds to the PEC condition \r{PEC}.
\item Denoting $M' =-\cot\TH/\h$, \r{THD'B'} yields $M'\#E_t+\#H_t=0$ which corresponds to the PEMC condition \r{PEMC}.
\end{itemize}

Since these properties of the DB and D'B' boundary do not depend on the $\#k$ vector of the incident plane-wave field (except that $k_x\not=0$), they are valid for any plane waves. Being linear conditions, they are equally valid for general fields which do not have sources at the boundary. It is interesting to notice that the DB and D'B' conditions appear complementary in showing PEC and PMC properties to TE and TM polarized fields. Moreover, the two PEMC admittances $M$ and $M'$ corresponding to the DB and D'B' boundary conditions satisfy the simple condition
\e MM' = -1/\h^2. \f
This result has a close connection to the duality transformation \cite{Methods}
\e \#E \ra \#E_d = j\h\#H,\ \ \ \ \ \#H\ra \#H_d = \#E/j\h, \f
which induces a transformation of media and boundary conditions. In particular, the DB and D'B' conditions are invariant in the transformation but the PEMC admittance is transformed as $M\ra M_d = -1/\h^2M=M'$.

\subsection{DB' and D'B boundaries}

Let us finally consider the two other boundary conditions \r{DB'} and \r{D'B} for the incident plane wave \r{Ei}, \r{Hi} satisfying the condition \r{TH}. For the DB' boundary the reflected fields satisfy
\e E_z^i+E_z^r=0,\ \ \ \ H_z^i-H_z^r=0. \f
From \r{Hr} the reflected transverse magnetic field component becomes
\e \h\#H^r_t = - \#u_x (k_z/k_x)\h H_z^i + \#u_y(k/k_x)E_z^i ,\f
which compared with \r{Hi} can be seen to equal $-\h\#H^i_t$. Because this is valid for any $\TH$ DB' boundary condition equals the PMC condition \r{PMC} for any fields except when $k_x=0$. 

Similarly, the condition \r{D'B} leads to  
\e E_z^i-E_z^r=0,\ \ \ \ H_z^i+H_z^r=0, \f
and from \r{Er} to
\e \#E^r_t = - \#u_x (k_z/k_x)E_z^i - \#u_y(k/k_x)\h H_z^i, \f
which when compared with \r{Ei} equals $-\#E_t^i$. This corresponds to the PEC condition \r{PEC}.

As a summary we can compare the four boundary conditions for TE and TM polarized incident fields in Table \ref{tab:BC}.

\begin{table}[ht]
	\centering
		\begin{tabular}{|c|c|c|}
		\hline
			 & TE$^z$ & TM$^z$\\
			 \hline
			 DB & PEC & PMC\\
			 D'B' & PMC & PEC\\ 
			 	 DB' & PMC & PMC\\
			 	 D'B & PEC & PEC\\
					 \hline			
		\end{tabular}
	\caption{Boundary conditions involving normal field components can be replaced by effective PEC and PMC conditions for fields with TE and TM polarizations.}
	\label{tab:BC}
\end{table}

\subsection{Reflection of TEM wave}

At this point one should consider the omitted case $k_x=0$ corresponding to the TEM wave incident normally to the boundary. Since in this case the incident wave does not have normal field components, the preceding conditions are already satisfied by the incident field and there appears to be no reflected field. However, since any reflected TEM wave can be added to the field, one may conclude that the conditions involving normal field components are not enough for defining the TEM fields. This anomaly, which was already pointed out by Rumsey \cite{Rumsey}, can be removed by adding the missing information. One may argue that a reflected TEM wave could also be associated to any TE and TM waves but, assuming a passive surface, it can be ruled out by power consideration. So the anomaly is associated with TEM waves, only, and we are free to define the reflection coefficient for the TEM wave (it may depend on the polarization of the wave).
 
The additional information comes naturally if the surface is taken as the limiting case of a given physical interface. For example, for the DB boundary studied in terms of its realization by a ZAP-medium interface \cite{DB,ZAP,EM10} the reflection of the TEM wave depends on the transverse medium parameters $\E_t,\M_t$ of the medium. For the special case when the wave impedance of the TEM wave is equal in both media, there is no reflection. In this case we can define the corresponding DB boundary by requiring no reflection for the TEM wave. Another ZAP medium leads to another definition. Similar consideration for the D'B'-boundary conditions requires a corresponding material realization which is yet to be found. For the DB' and D'B boundaries it appears quite natural to assume that the respective PMC and PEC conditions are valid for the TEM wave as well.

For a localized source giving rise to a continuous spectrum of plane waves, the normally incident component has zero measure, i.e., it corresponds to zero portion of the total radiated power. Thus, the TEM component plays no role and can be neglected, as was already pointed out by Rumsey \cite{Rumsey}. Also, in the case of a non-planar boundary an incident plane wave has a component appearing as a local TEM wave with zero energy which also can be omitted. It was shown through numerical computations for plane-wave scattering from spherical and cubic scatterers that there is no back-scattered wave when the DB or D'B' conditions are satisfied at the scatterer as predicted by the theory \cite{sphere,EM9}.

\subsection{Other possibilities}

For completeness, the two remaining possible combinations of boundary conditions, 
\e D_z=0,\ \ \ \ \ \d_zD_z=0, \l{1}\f
\e B_z=0,\ \ \ \ \ \d_zB_z=0, \l{2}\f
appear to be of little use. As an example, imposing \r{1} on the previous plane wave yields 
\e E_z^i+E_z^r=0,\ \ \ E_z^i-E_z^r=0,\ \ \ \ \ \Ra\ \ \ \ \ E_z^i=E_z^r=0. \f 
This restricts the freedom of choice of the incident field. Thus, launching a TM incident field creates a contradiction, which can be so interpreted that \r{1} is of improper form.

\section{More general boundary surfaces}

Let us generalize the previous analysis by replacing the planar boundary surface by one defined by a function $x_3(\#r)$ as $x_3(\#r)=0$. Defining two other functions $x_1(\#r)$ and $x_2(\#r)$ so that they make a system of orthogonal coordinates satisfying 
\e \D x_i(\#r)\.\D x_j(\#r)=0,\ \ \ \ i\not=j, \f
allows us to express various differential operators in the form given in Appendix 2.  

Expressing the fields as
\e \#E = \sum_{i=1}^3 \#u_i E_i, \ \ \ \ \  \#H = \sum_{i=1}^3 \#u_i H_i, \f
and similarly for the $\#D$ and $\#B$ fields, the DB-boundary conditions are expressed by
\e D_3=0,\ \ \ \ B_3=0, \f
for $x_3=0$. The conditions for the D'B' boundary are somewhat more complicated. Writing the expansion of the divergence
\e \D\.\#F = \D_t\.\#F_t + \frac{1}{h_1h_2h_3}\d_{x_3}(h_1h_2F_3), \f 
for the Cartesian coordinates with $x_3=z$ as 
\e \D\.\#F = \D_t\.\#F_t + \d_zF_z, \f
gives us a reason to anticipate that a boundary condition of the form $\d_zF_z=0$ for the planar surface should take the form $\d_{x_3}(h_1h_2F_3)=0$ for the more general surface.

\subsection{Boundary conditions}

Assuming an isotropic medium bounded by the surface $S: x_3=0$, we can propose four possible boundary conditions on $S$ involving only the normal field components $E_3$ and $H_3$ and their normal derivatives as
\e E_3=0\ \ \ {\rm and}\ \ \ H_3=0, \l{EH3}\f
\e E_3=0\ \ \ {\rm and}\ \ \ \d_{x_3}(h_1h_2H_3)=0, \l{EdH3}\f
\e \d_{x_3}(h_1h_2E_3)=0\ \ \ {\rm and}\ \ \ H_3=0, \l{dEH3}\f
\e \d_{x_3}(h_1h_2E_3)=0\ \ \ {\rm and}\ \ \ \d_{x_3}(h_1h_2H_3)=0. \l{dEdH3}\f

Let us first consider consequences of these conditions. For $E_3=0$ on the boundary $S$ one of the Maxwell equations yields  
\e \#u_3\.(\D\x\#H) = \frac{1}{h_1h_2}(\d_{x_1}(h_2H_2)- \d_{x_2}(h_1H_1))=0, \l{E30}\f
while for $H_3 = 0$ the other Maxwell equation yields
\e \#u_3\.(\D\x\#E) = \frac{1}{h_1h_2}(\d_{x_1}(h_2E_2)- \d_{x_2}(h_1E_1))=0. \l{H30}\f
\r{E30} is satisfied if there exists a scalar function $\psi(x_1,x_2)$ such that on $S$ we can write 
\e H_1(x_1,x_2)  = \frac{1}{h_1}\d_{x_1}\psi(x_1,x_2),\f  
\e H_2(x_1,x_2)  = \frac{1}{h_2}\d_{x_2}\psi(x_1,x_2). \f
Similarly, \r{H30} is satisfied for a function $\phi(x_1,x_2)$ and
\e E_1(x_1,x_2)  = \frac{1}{h_1}\d_{x_1}\phi(x_1,x_2),\f
\e E_2(x_1,x_2)  = \frac{1}{h_2}\d_{x_2}\phi(x_1,x_2), \f
as can be easily checked. Thus, \r{E30} and \r{H30} can be compactly expressed in vector form as
\e E_3=0\ \ \ \ \Ra\ \ \ \ \#H_t(x_1,x_2) = \D_t\psi(x_1,x_2), \l{E3}\f
\e H_3=0,\ \ \ \ \Ra\ \ \ \ \#E_t(x_1,x_2) = \D_t\phi(x_1,x_2). \l{H3}\f
On the other hand, outside of sources, the divergence of the fields vanishes, whence from \r{divt} we can write
\e \d_{x_3}(h_1h_2E_3(\#r))=0,\ \ \ \ \Ra\ \ \ \ \ \D_t\.\#E_t(\#r) =0, \l{DEr3}\f
\e \d_{x_3}(h_1h_2H_3(\#r))=0,\ \ \ \ \Ra\ \ \ \ \ \D_t\.\#H_t(\#r) =0. \l{DHr3}\f
Let us apply these on the different combinations of boundary conditions \r{EH3} -- \r{dEdH3}.

\subsection{DB boundary}

Starting from \r{EH3} $(E_3=0, H_3=0)$ the fields on $S$ can be expressed in terms of two potential functions as given in \r{E3}, \r{H3}. Let us now consider two special cases. Fields satisfying $E_3(\#r)=0$ everywhere will be called TE$^3$ fields and fields satisfying $H_3(\#r)=0$ will be called TM$^3$ fields. 

Obviously, a TE$^3$ field satisfies $\d_3(h_1h_2E_3)=0$ everywhere, including the boundary surface $S$. From \r{H3} and \r{DEr3} we conclude that on $S$ the potential $\phi$ satisfies
\e \D_t\.\#E_t = \D_t\.(\D_t\phi(x_1,x_2))= \D_t^2\phi(x_1,x_2)=0. \f
From the discussion in the Appendix we conclude that this implies
\e \#E_t = \D_t\phi(x_1,x_2)=0, \f
i.e., PEC condition on the boundary surface $S$. The result can be generalized to problems where $S$ is not closed but extends to infinity, provided the sources are localized so that the fields vanish in the infinity.

Similarly, for the TM$^3$ field the potential $\psi$ must satisfy
\e \D_t^2\psi(x_1,x_2)=0 \l{D2psi}\f
on the surface $S$, whence the TM$^3$ field sees the boundary defined by the DB conditions \r{EH3} as a PMC boundary. This condition can also be obtained from the duality transformation which swaps electric and magnetic fields and, hence, PEC and PMC conditions. The DB boundary is invariant to the duaity transformation. 

It must be pointed out that there is no guarantee that a given field can be expressed as a sum of partial fields TE and TM polarized with respect to the coordinate function $x_3(\#r)$ in the general case. Such a decomposition is known to be valid with respect to Euclidean coordinates and the radial spherical coordinate.

\subsection{D'B' boundary}

Considering now the boundary conditions \r{dEdH3}, from \r{DEr3}, \r{DHr3} the fields must satisfy 
\e \D_t\.\#E_t(x_1,x_2)=0,\ \ \ \ \D_t\.\#H_t(x_1,x_2)=0 \l{DtEH}\f 
at the boundary. From \r{E3} a TE$^3$ field can be represented in terms of a potential $\psi(x_1,x_2)$ which from \r{DtEH} satisfies \r{D2psi}. From \r{int} we again conclude that a TM$^3$ field sees a D'B' boundary as a PMC boundary. Similarly a TM$^3$ field sees the same boundary as a PEC boundary. Thus, in this respect DB and D'B' boundaries show complementary properties. 

It is clear that the D'B' conditions \r{dEdH3} represent a generalization of the conditions \r{EzHz} for boundary surfaces more general than the planar surface. Comparing \r{dEdH3} and \r{divt} whose last term can be written as $\D\.(\#u_3\#u_3\.\#F)$, we see that the proper form for the D'B'-boundary conditions is
\e \D\.(\#n\#n\.\#E)=0,\ \ \ \ \ \D\.(\#n\#n\.\#H)=0. \f

As an example, for the spherical coordinates $x_1=\TH,\ x_2=\VF,\ x_3=r$ the metric coefficients are 
\e h_1=h_\TH = r,\ \ \ h_2=h_\VF=r\sin\TH,\ \ \ h_3=h_r=1, \f
and the D'B'-boundary conditions \r{dEdH3} at the surface $x_3=r=a$ have the form
\e \d_r(r^2E_r)=0,\ \ \ \ \  \d_r(r^2H_r)=0 \f 
instead of $\d_rE_r=0, \d_rH_r=0$ as would be suggested by \r{EzHz}.

\subsection{DB' and D'B boundaries}

The mixed conditions \r{EdH3} and \r{dEH3} can be handled in the same  way. In the case \r{EdH3}, \r{E30} implies $\#H_t=\D_t\psi$ on $S$ while \r{DHr3} implies $\D_t^2\psi=0$. From the reasoning given in the Appendix we obtain $\#H_t=0$ on $S$. Thus, \r{EdH3} corresponds to the PMC conditions for any fields. Similarly, we can show that \r{dEH3} corresponds to the PEC conditions for any fields.

\subsection{The TF$^3$ field}

The DB and D'B' conditions were tested above for the TE$^3$ and TM$^3$ fields. Let us finally consider their generalization in terms of a combined field 
\e \#F=\sin\TH \#E + \cos\TH\h \#H, \f
where $\TH$ is a parameter. Its component 3 is assumed to satisfy everywhere the condition
\e F_3=\sin\TH E_3 + \cos\TH\h H_3=0. \l{F3}\f
Any field satisfying \r{F3} is called a TF$^3$ field. Since $\d_3(h_1h_2F_3)=0$ everywhere, from \r{divt} the TF$^3$ field satisfies 
\e \D_t\.\#F_t= \D_t\.(\sin\TH \#E_t + \cos\TH\h \#H_t)=0. \l{div1}\f
Inserting the Maxwell equations we can expand
\e \#F = \frac{1}{jk}\D\x(\sin\TH \h\#H - \cos\TH \#E), \f 
whence the TF$^3$ field satisfies 
\e \#u_3\.\D_t\x(\sin\TH\h\#H_t - \cos\TH \#E_t)=0. \l{curl1}\f
Both \r{div1} and \r{curl1} are valid in source-free regions. 

Because at the DB boundary $\#E_t$ and $\#H_t$ are curl-free, \r{curl1} is automatically valid. The DB condition \r{EH3} implies
\e \#u_3\.\D_t\x(\sin\TH\#E_t + \cos\TH \h\#H_t)=0 \l{curl2}\f
at the boundary. This combined with \r{div1} and the reasoning for a closed surface given in the Appendix leads to
\e \sin\TH\#E_t + \cos\TH \h\#H_t=0 \f
at the DB boundary. This equals the PEMC condition 
\e M\#E_t + \h\#H_t=0,\ \ \ \ \ M = \tan\TH/\h. \f 
 
Following a similar path of reasoning, one can see that at the D'B' boundary \r{div1} is automatically valid for a TF$^3$ field while \r{curl1} and 
\e \D_t\.(\sin\TH \h\#H_t - \cos\TH\#E_t)=0, \l{div2}\f
which follows from the D'B' conditions \r{dEdH3}, correspond to the PEMC condition 
\e M'\#E_t + \h\#H_t=0,\ \ \ \ \ M' = -\cot\TH/\h. \f 

As a conclusion, for a TF$^3$ field, which is a generalization of the TE$^3$ and TM$^3$ fields, both DB and D'B' boundaries appear as PEMC boundaries with the respective admittances $M$ and $M'$. This includes as special cases the results for the TE$^3$ and TM$^3$ fields given above.

\section{Discussion}

The four boundary conditions \r{EH3} -- \r{dEdH3} involving only field components normal to the boundary surface and their normal derivatives, form an interesting set. Two of these, \r{EdH3} and \r{dEH3} are alternatives for the respective PMC and PEC conditions which in terms of tangential fields are normally taken in the form \r{PMC} and \r{PEC}. The other two conditions, dubbed as DB and D'B' conditions, \r{EH3}, \r{dEdH3}, appear new as kind of mixtures of PEC and PMC conditions or variants of the PEMC condition. The DB conditions have been introduced already in 1959 \cite{Rumsey} but have not been applied for half a century; the D'B' conditions have not been discussed earlier to the knowledge of these authors. Since their representations are so basic, DB and D'B' conditions seem to have their right of existence along with the PEC and PMC conditions.

In introducing new boundary conditions the first question is to find their properties for the electromagnetic fields. This has been done here and in some of the previous publications. The next step would be to ponder about their possible applications. In case there are some of enough interest, the problem of practical realization of such boundary surfaces as an interface of some material comes up. As explained in the introduction, the planar DB boundary may have some application as an isotropic soft surface. Also, as shown in \cite{Kong,Yaghjian}, the DB conditions play a central role in the theory of electromagnetic cloaking. The DB boundary can be realized, e.g.,  by an interface of a uniaxial anisotropic medium with zero axial parameters (ZAP medium) or another medium called the uniaxial IB medium  \cite{IBDB,ZAP}. Finding a realization for the D'B' medium is still an open problem.

In \cite{EM9} it has been shown that both DB and D'B' boundaries are self dual which implies that scatterers symmetric in $\pi/2$ rotation have no backscattering. This is an interesting property, shown to be valid by numerical computations, and it may have potential applications. As a continuation of the present study, it has been further shown that it is possible to define a generalized DB boundary conditions of which the DB and D'B' conditions are special cases. Such boundaries are also self dual and have no backscattering \cite{EM10}.

In addition to the interesting physical properties of the novel boundary conditions, there may be some advantage in numerical electromagnetics by considering normal field components on the surface instead of the tangential components in the cases of DB' and D'B boundaries which correspond to the respective PMC and PEC conditions for the tangential fields.

\section*{Appendix 1: Fields in uniaxial IB medium}

A medium called uniaxial IB medium has been defined in \cite{IBDB} by conditions of the form
\e \#D = a\#B_t + b\#u_zB_z + e\#u_z\x\#E, \l{D}\f
\e \#H = m\#u_z\x\#B + c\#E_t + d\#u_zE_z, \l{H}\f
in terms of six parameters $a,b,c,d,e,m$. Let us briefly consider fields in such a medium. Inserting \r{D} and \r{H} in the second one of the source-free Maxwell equations
\e \D\x\#E = -j\o\#B, \l{MaxE} \f
\e \D\x\#H = j\o\#D, \l{MaxH} \f
we can split both of them in two components. The axial components are, respectively,
\e \#u_z\.\D_t\x\#E_t = -j\o B_z, \l{ax1}\f
\e m\D_t\.\#B_t + c\#u_z\.\D_t\x\#E_t = j\o b B_z \l{ax2}. \f
Eliminating the $B_z$ component yields the relation
\e (b+c)\#u_z\.\D_t\x\#E_t = -m\D_t\.\#B_t \l{rel1} \f
between the transverse fields. The transverse components of \r{MaxE} and \r{MaxH} can be written as
\e \d_z\#u_z\x\#E_t + j\o\#B_t = \#u_z\x\D_t E_z, \l{tr1} \f
\e (c\d_z - j\o e)\#u_z\x\#E_t - (m\d_z+j\o a)\#B_t = d\#u_z\x\D_t E_z. \l{tr2} \f
Taking the divergence of \r{tr1} and \r{tr2} yields
\e \d_z\#u_z\.\D_t\x\#E_t =j\o\D_t\.\#B_t, \l{rel2} \f
\e (c\d_z - j\o e)\#u_z\.\D_t\x\#E_t = -(m\d_z+j\o a)\D_t\.\#B_t. \l{rel3} \f

The three equations \r{rel1}, \r{rel2} and \r{rel3} can now be reduced as follows. First we eliminate the term $\#u_z\.\D_t\x\#E_t$ leaving us with the two equations
\e [m\d_z + j\o(b+c)]\D_t\.\#B_t =0, \l{eq1} \f
\e [mb\d_z + j\o(me + a(b+c))]\D_t\.\#B_t=0. \l{eq2} \f
As a second step we eliminate $\d_t(\D_t\.\#B_t)$ and arrive at the single equation
\e [(b+c)(b-a)-me]\D_t\.\#B_t=0. \f
Omitting the special case when the factor in brackets vanishes which corresponds to a certain subset of the medium, we conclude that the transverse $\#B$ vector must be divergenceless, 
\e \D_t\.\#B_t=0. \l{DBt}\f
From \r{rel1} it then follows that the $\#E$ vector must satisfy
\e \#u_z\.\D_t\x\#E_t=0 \l{uDEt}\f
everywhere in the medium. Here we have also assumed that $b+c\not=0$. Finally, from \r{ax1} and \r{D} we obtain conditions for the axial field components,
\e B_z=0, \ \ \ \ \ D_z=0 \l{BzDz}. \f

To conclude, we have shown that any field in the uniaxial IB medium, defined by the medium equations \r{D}, \r{H} in their general form, must satisfy the conditions \r{BzDz}. Thus, if we cut a layer, however thin, of the medium with two planes orthogonal to the $z$ axis, the fields on both sides of the slab must satisfy the DB-boundary conditions \r{DB} with $\#n=\#u_z$.

\section*{Appendix 2: Orthogonal coordinates}

The following review of differential operators for orthogonal coordinates defined by three functions $x_1(\#r), x_2(\#r)$ and $x_3(\#r)$ has been taken from \cite{Bladel,Jeffrey}. The boundary surface $S$ is defined by $x_3(\#r)=0$. In this case, $x_1(\#r)$ and $x_2(\#r)$ define orthogonal coordinates on $S$. The coordinate unit vectors can be represented as
\e \#u_i = h_i \D x_i(\#r), \ \ \ \ \ h_i = (\D x_i(\#r)\.\D x_i(\#r))^{-1/2}, \l{ui}\f
where the $h_i=h_i(\#r)$ are the metric-coefficient functions. These follow from the definition of the gradient,
\e \D f(\#r) = \sum_{i=1}^3 \frac{\#u_i}{h_i}\d_{x_i}f(\#r). \l{grad}\f
The divergence and curl of a vector function
\e \#F(\#r) = \sum_{i=1}^3 \#u_iF_i(\#r) \f
are defined by
\e \D\.\#F = \frac{1}{h_1h_2h_3}\sum_{i=1}^3 \d_{x_i}\le\frac{h_1h_2h_3}{h_i}F_i\ri, \l{div} \f
$$ \D\x\#F = \frac{\#u_1}{h_2h_3}(\d_{x_2}(h_3F_3)- \d_{x_3}(h_2F_2)) $$
$$ +\frac{\#u_2}{h_3h_1}(\d_{x_3}(h_1F_1)- \d_{x_1}(h_3F_3)) $$ 
\e +\frac{\#u_3}{h_1h_2}(\d_{x_1}(h_2F_2)- \d_{x_2}(h_1F_1)), \l{curl}\f
and the Laplacian by
\e \D^2 f = \D\.(\D f) = \frac{1}{h_1h_2h_3}\sum_{i=1}^3 \d_{x_i}\le\frac{h_1h_2h_3}{h_i^2}\d_{x_i}f\ri. \l{lap} \f

The differential operations can be split in parts along the special coordinate $x_3$ and transverse to it:
\e \D f = \D_t f + \frac{\#u_3}{h_3}\d_3 f, \f
\e \D\.\#F = \D_t\.\#F_t + \frac{1}{h_1h_2h_3}\d_{x_3}(h_1h_2F_3), \l{divt}\f 
\e \D\x\#F = (\D\x\#F)_t + \frac{\#u_3}{h_1h_2}(\d_{x_1}(h_2F_2)- \d_{x_2}(h_1F_1)). \f
\e \D^2 f = \D_t^2 f + \frac{1}{h_1h_2h_3} \d_{x_3}\le\frac{h_1h_2}{h_3}\d_{x_3}f\ri.  \f

If a scalar function $\phi(x_1,x_2)$ satisfies on the surface $S: x_3=0$ the Laplace equation
\e \D_t^2\phi(x_1,x_2) =0, \l{D2phi}\f
and if the surface is closed, we can expand the surface integral as
\e \I_S |\D_t\phi|^2dS = \I_S \D_t\.(\phi\*\D_t\phi)dS - \I_S \phi\*\D_t^2\phi dS.  \l{int}\f
Both integrals on the right-hand side vanish. The last one because \r{D2phi} and the middle one because the surface is closed. Thus, we can conclude that $\D_t\phi=0$ on $S$. This elementary proof is, however, valid for simply connected surfaces only, and requires a more involved analysis for more complicated surfaces like the torus, see \cite{Kress,Gulzow}. 

This condition remains valid for an open surface $S$ extending to infinity when the integral over the boundary contour $C$ of $S$ in infinity,
\e \I_S \D_t\.(\phi\*\D_t\phi)dS = \OI_C \#m\.(\phi\*\D_t\phi)dC, \l{IS}\f
vanishes. Here $\#m$ is the unit vector normal to the contour and parallel to the surface in infinity. For a planar surface $z=0$ we have $\#m=\#u_\R$, the radial unit vector. As an example, if $\D_t\phi$ is the tangential component of the electric field from a localized source, on the contour integral $\#m\.\D_t\phi$ corresponds to the radial component of the far field which is known to decay as $1/\R^2$ along the plane. Since $dC=\R d\VF$, the integral vanishes and $\D_t\phi=0$ on the plane which corresponds to the PEC condition. Hovever, this is not necessarily the case when the source extends to infinity. For example, for the normally incident TEM plane wave with constant $\#E_t$ we have $\phi=\#E_t\.\RR$, whence the integral \r{IS} becomes infinite. A more complete analysis of the open surface case remains still to be done. 

\section*{Acknowledgment}
This work has been partly supported by the Academy of Finland. Discussions with Professor P.-S. Kildal on the anomaly appearing for normally incident plane waves are gratefully acknowledged.

\end{document}